\documentclass[10pt,letterpaper,compsoc,conference]{iiswc26}

\usepackage{cite}
\usepackage{algorithmic}
\usepackage{graphicx}
\usepackage[dvipsnames]{xcolor}
\usepackage[final]{microtype}
\usepackage[italic]{mathastext}
\usepackage[T1]{fontenc}
\usepackage{orcidlink}
\usepackage{textcomp}
\usepackage{underscore}
\usepackage{float}
\usepackage{booktabs}
\usepackage[varqu,varl]{zi4}
\usepackage[all]{nowidow}
\usepackage[auth-lg,affil-it]{authblk}
\usepackage[keeplastbox]{flushend}
\usepackage{fancyhdr}
\usepackage{xspace}
\PassOptionsToPackage{hyphens}{url}
\usepackage{hyperref}

\usepackage{tikz}
\usetikzlibrary{positioning}
\usetikzlibrary{arrows.meta}
\usepackage{subcaption}

\newcommand{\Erdos}{Erd\H{o}s-R\'enyi\xspace}

\begin{document}

\title{ChainForge: Characterizing Embedding as the Bottleneck in Quantum Annealer Workloads}





 \author{Kanishka Jayathilake\orcidlink{0009-0005-8751-9554}}
\author{Cordelia Brumley\orcidlink{0009-0001-6141-6785}}
 \author{Tanner Smith\orcidlink{0009-0003-8630-9884}}
 \author{Ramin Ayanzadeh\orcidlink{0000-0001-6687-5668}}
 \affil{University of Colorado Boulder}


\maketitle

\pagestyle{plain}


\begin{abstract}
Quantum Annealers (QAs) are among the first commercially scaled quantum computing systems designed for large-scale optimization.
Unlike digital systems that execute sequences of compiled instructions, QAs operate as analog single-instruction machines that directly evolve an Ising Hamiltonian toward low-energy solutions.
To execute an application, the logical problem graph must first be mapped onto the hardware's sparse connectivity via embedding, where logical variables are represented by chains of connected physical qubits.
As a result, embedding becomes the dominant system challenge in QAs, shaping whether and how workloads can execute on the machine.
Despite its central role, embedding has largely been treated as a preprocessing step rather than a system bottleneck.

In this work, we present \emph{ChainForge}, the first systems and architecture characterization of embedding in QA workloads.
Using diverse workload families and graph topologies on modern QA hardware, we characterize how embedding impacts effective hardware capacity, routing overheads, runtime variability, and solution quality.
Our results show that embedding inflates physical resource usage, long chains degrade annealing fidelity and scalability, and heuristic embedders may fail even when valid embeddings exist.
We further show that embedding latency can become a runtime bottleneck for dynamic workloads requiring frequent remapping, while nominal qubit counts significantly overestimate the usable capacity of QAs for realistic applications.
Overall, our findings establish embedding as the defining workload bottleneck and systems abstraction of QAs, providing architectural insights for future hardware topologies, runtime systems, and workload-aware annealing platforms.

\end{abstract}

\section{Introduction}

Quantum computing is emerging as a promising paradigm for solving classes of problems that remain beyond the practical reach of classical supercomputers \cite{preskill2018quantum, feynman1982simulating,NielsenChuang2010,guang2025potential,papadopoulos2026stabilizer}.
Quantum systems broadly fall into two categories: digital quantum systems, which execute workloads through quantum circuits \cite{shor1994algorithms,grover1996fast,NielsenChuang2010}, and analog quantum systems, which evolve a physical system toward a desired solution~\cite{albash2018adiabatic,ayanzadeh2020leveraging,ayanzadeh2022equal}.
Unlike classical computing, where analog systems are typically specialized, both digital and analog quantum systems can support universal computation in theory \cite{albash2018adiabatic,aharonov2008adiabatic}.
Among analog quantum systems, Quantum Annealers (QAs) have emerged as one of the first commercially scaled quantum computing platforms \cite{D-Wave, mcgeoch2021advantage} and are increasingly being explored for large-scale optimization workloads across domains including scheduling, logistics, finance, machine learning, cryptography, and drug discovery \cite{Yarkoni_2022, phillipson2021portfolio,  venturelli2015quantum, inoue2021traffic, elsokkary2017financial, o2018nonnegative, hu2020quantum,mulligan2020designing,ayanzadeh2019quantum,ayanzadeh2020ensemble,ayanzadeh2018solving}.
As QA systems continue to scale and become accessible through commercial cloud platforms, understanding their architectural behavior and workload limitations is becoming increasingly important for the systems and architecture community.

\begin{figure}[b]
    \centering
    \includegraphics[width=\columnwidth]{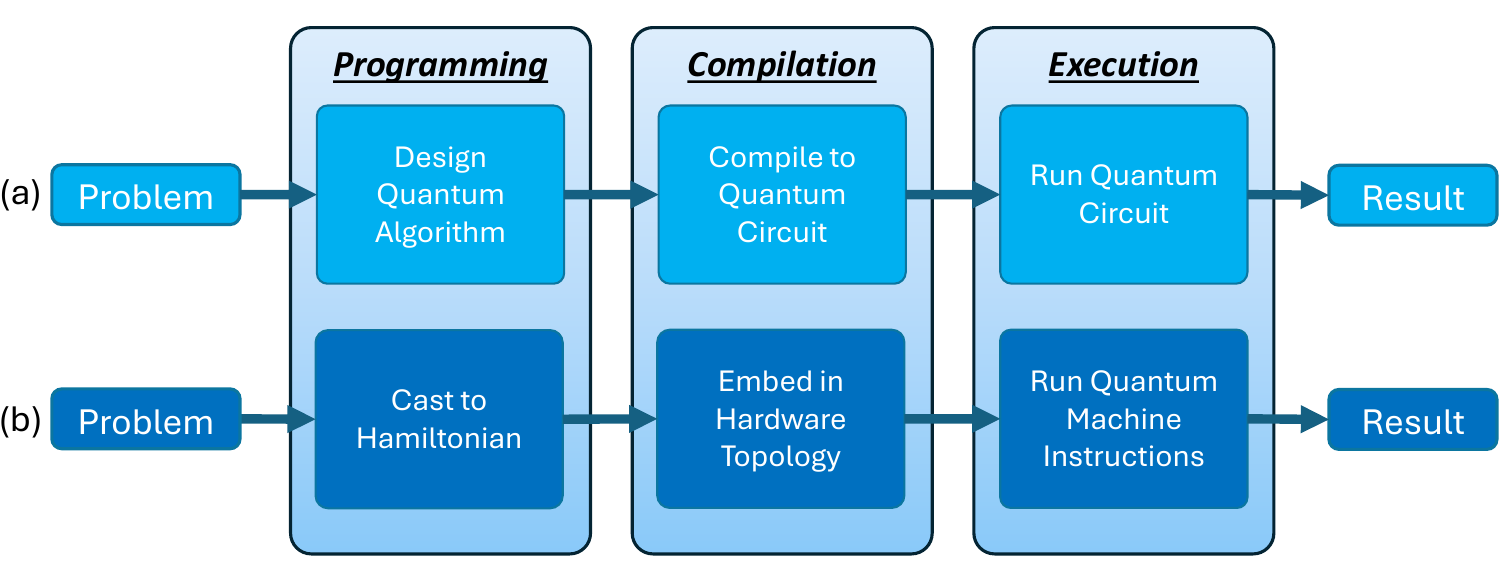}
    \caption{
        Execution models of (a) digital quantum computers and (b) analog quantum annealers, showing their main stages.
    }
    \label{fig:qc_models}
\end{figure}

The execution model of QAs fundamentally differs from that of digital quantum systems (Fig.~\ref{fig:qc_models}).
Digital quantum systems execute workloads as sequences of quantum operations that are compiled into quantum circuits and dynamically mapped, routed, and scheduled across physical qubits during execution~\cite{murali2019noise,tannu2019ensemble,tannu2018case,molavi2026generating}.
QAs instead operate as analog single-instruction machines that evolve a physical system, called an \emph{Ising Hamiltonian}, toward low-energy states representing candidate solutions \cite{kadowaki1998quantum,ayanzadeh2023enhancing,ayanzadeh2022quantum,ayanzadeh2022equal}.
This execution model eliminates instruction scheduling during execution, but introduces a fundamental architectural constraint: all connectivity requirements must be resolved before execution begins~\cite{choi2008minorembeddingadiabaticquantumcomputation,ayanzadeh2023enhancing,ayanzadeh2022equal,ayanzadeh2024skipper}.
Consequently, the scalability and execution behavior of QAs are fundamentally shaped by how workloads are mapped onto hardware connectivity.

Embedding is the process of mapping the problem graph onto the physical connectivity graph of a QA.
Because the problem and hardware topologies generally differ, embedding may represent a logical variable using a chain of connected physical qubits, with strong couplings enforcing the chain to act as a single variable during annealing.
These chains enable the logical interactions required by the problem to be realized using the sparse physical couplers available in hardware.
Unlike digital quantum systems, which can dynamically route interactions using operations such as SWAPs~\cite{tannu2019ensemble,ayanzadeh2023frozenqubits,ayanzadeh2023enigma,murali2019noise}, QAs operate as single-instruction machines and must resolve these connectivity constraints through embedding before annealing begins.

From a systems and architecture perspective, embedding fundamentally defines the effective capacity of QAs.
Although modern QAs contain thousands of physical qubits \cite{boothby2020Pegasus, boothby2021zephyr}, embedding overheads can reduce fully connected workloads to only around 177 logical variables~\cite{ayanzadeh2024skipper}.
Thus, workload topology, hardware connectivity, and embedding jointly determine the practical scalability and execution fidelity of QAs \cite{lobe2021minor,ayanzadeh2024skipper,ayanzadeh2023enhancing}.

Embedding is also a computational bottleneck.
Finding an optimal embedding is NP-hard, forcing practical QA systems to rely on heuristic algorithms that may require substantial runtime and can fail even when feasible embeddings exist \cite{gomeztejedor2026addressingminorembeddingproblemquantum,bernal2020integer}.
Embedding latency and reliability therefore become critical for dynamic workloads that require repeated remapping as their structure changes.

Despite its central role, embedding has largely been studied from the perspective of embedding algorithms, parameter tuning, or workload-specific optimizations.
Prior work has explored embedding heuristics, chain-strength tuning, annealing schedules, application-specific mappings, and benchmarking of embedding algorithms \cite{gomeztejedor2026addressingminorembeddingproblemquantum, macaskillsmith2026emberextensiblebenchmarksuite, Raymond_2020, klymko2014adiabatic, zbinden2020embedding, goodrich2018optimizing, barbosa2021optimizing}.
However, the systems and architecture community still lacks a workload-centric characterization of how embedding impacts effective hardware capacity, routing behavior, workload scalability, runtime variability, and annealing fidelity across diverse workload families and graph topologies.
This leaves fundamental architectural questions unanswered regarding how workload structure interacts with hardware topology and how embedding behavior shapes practical QA scalability and execution fidelity.

To address this gap, we present \emph{ChainForge}, the first systems and architecture characterization study of embedding in QA workloads.
ChainForge systematically analyzes embedding behavior across diverse workload families and modern QA hardware topologies.
Our study characterizes effective hardware capacity loss caused by physical-qubit inflation and routing fragmentation, chain structures and routing overheads across workload topologies, embedding runtime and heuristic variability, and the relationship between embedding quality and annealing fidelity.
Our characterization reveals that nominal qubit counts substantially overestimate usable QA capacity for realistic workloads, while dense and irregular graph topologies rapidly amplify embedding overheads, and heuristic embedders may fail despite the existence of feasible mappings.
More importantly, we show that embedding quality fundamentally shapes QA solution quality and execution reliability.
We further show that embedding latency can become a bottleneck for dynamic workloads requiring repeated remapping.
Our findings establish embedding as the defining workload bottleneck and critical systems abstraction of QAs, providing architectural insights for future  annealing hardware, runtime, and workload aware embedding frameworks.

\vspace{0.5in}
\noindent
Overall, this work makes the following contributions:

\noindent\hangindent=1em\hangafter=0 $\bullet$
We present \emph{ChainForge}, the first systems and architecture characterization study of the embedding problem in quantum annealer workloads.

\noindent\hangindent=1em\hangafter=0 $\bullet$
We characterize the impact of embedding on effective QA capacity, showing that sparse hardware connectivity and chain formation substantially inflate physical-qubit usage and reduce usable logical capacity.

\noindent\hangindent=1em\hangafter=0 $\bullet$
We analyze how workload topology shapes embedding behavior across diverse graph families, characterizing chain structures, routing fragmentation, embedding variability, and scalability bottlenecks.

\noindent\hangindent=1em\hangafter=0 $\bullet$
We study the computational behavior of embedding heuristics, showing that embedding itself becomes a runtime bottleneck due to high latency, heuristic variability, and failures even when feasible embeddings exist.

\noindent\hangindent=1em\hangafter=0 $\bullet$
We characterize the impact of embedding quality on QA execution fidelity, showing that embedding behavior directly affects chain-break rates, execution reliability, and annealing success.

\begin{figure*}[h]    
    \centering    
    \captionsetup[subfigure]{position=top} 
    \centering
    \subfloat[]{
        \includegraphics[width=0.58\textwidth]{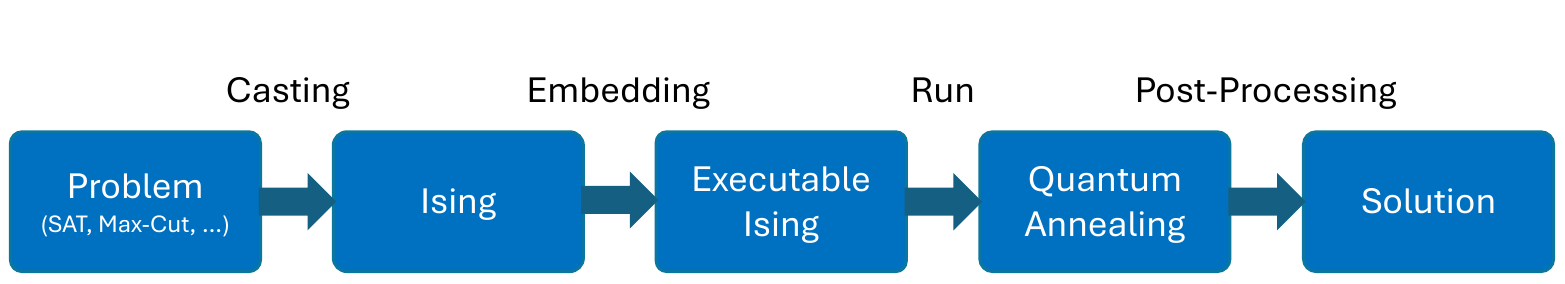}
    }
    \subfloat[]{        
        \includegraphics[width=0.4\textwidth]{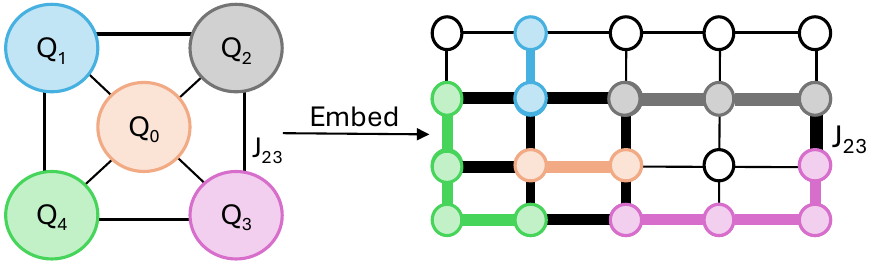}
    }
    \caption{          
        (a) Stages of the quantum annealing pipeline.  
        (b) Example embedding a problem graph onto a QA hardware graph.
    }    
    \label{fig:background} 
\end{figure*}

\section{Background and Motivation}
Executing workloads on QAs requires transforming abstract optimization problems into hardware-compatible Ising programs.
This section provides the technical background required to understand QA execution and the role of embedding in QA workloads.

\subsection{Quantum Annealer: An Analog Quantum Accelerator for Optimization}
QAs are analog quantum accelerators designed for optimization workloads.
From a systems perspective, QAs operate as single-instruction machines that sample low-energy states of the following Ising Hamiltonian \cite{farhi2001quantum}:
\begin{equation}
    E(s)=\sum_i h_i s_i + \sum_{i<j} J_{ij} s_i s_j,
    \label{eq:ising}
\end{equation}
where $s_i \in \{-1,+1\}$ denotes the spin state of variable $i$, $h_i$ represents local biases, and $J_{ij}$ represents pairwise couplings between variables.
In this regime, with tolerable noise, lower-energy configurations are exponentially more likely to be sampled than higher-energy ones~\cite{AyanzadehMultiQubit}.

\subsection{QA Workflow}
Figure~\ref{fig:background}(a) illustrates the end-to-end QA execution workflow.
As QAs operate as single-instruction optimization accelerators, applications must first be transformed into an Ising formulation before execution.
This process, referred to as \emph{casting}, maps the original optimization problem into an Ising Hamiltonian whose minimum-energy state corresponds to the desired solution~\cite{McGeoch2020169,lucas2014ising, Ayanzadeh2020Reinforcement}.
Many applications, including graph coloring, TSP, SAT, scheduling, routing, and portfolio optimization, can be naturally expressed in this representation \cite{Yarkoni_2022}, yielding a logical Ising graph in which vertices represent logical variables and edges encode pairwise interactions, thereby capturing the structure and connectivity of the original optimization problem.

However, as physical qubits are sparsely connected, logical Ising graphs cannot typically be executed directly on QA hardware \cite{boothby2020Pegasus, boothby2021zephyr}.
Embedding transforms the logical graph into a hardware-compatible representation supported by the target QA architecture \cite{choi2008minorembeddingadiabaticquantumcomputation}.
The resulting embedded graph is then converted into a hardware-executable representation by truncating and scaling coefficients to match hardware precision constraints and by specifying execution control parameters such as the annealing schedule.
After annealing, measured outputs are post-processed to recover the solution of the original optimization problem.

\subsection{Embedding: A Bottleneck in QA Workloads}
Embedding maps logical Ising graphs onto sparse QA hardware topologies.
Since practical optimization workloads often require denser interactions than what is supported by the hardware graph, a logical variable cannot always be mapped onto a single physical qubit.
To address this mismatch, logical variables are mapped onto groups of connected physical qubits, called \emph{chains}~\cite{barbosa2021optimizing}.

A chain represents a single logical variable using multiple physical qubits distributed across the hardware graph (Fig.~\ref{fig:background}(b)).
The qubits belonging to a chain must form a connected component such that they collectively behave as the same logical variable during annealing \cite{choi2008minorembeddingadiabaticquantumcomputation}.
Additionally, if two logical variables interact in the original graph, at least one physical coupler must connect the corresponding physical chains on the hardware graph.
Embedding also transforms logical Ising coefficients into hardware-level parameters, ensuring that the global minimum of the Hamiltonian executed on QA hardware corresponds to the global minimum of the original logical Hamiltonian.

Physical qubits within a chain are strongly coupled using ferromagnetic interactions such that they converge to the same value during annealing \cite{Raymond_2020}.
This coupling magnitude is controlled through a parameter called \emph{chain strength}, which acts as a consistency mechanism for maintaining logical coherence across the chain \cite{dwave_chain_strength_intro, dwaveocean, ayanzadeh2022equal, ayanzadeh2024skipper}.

\subsection{Architectural Implications of Embedding}
Embedding introduces substantial hardware, runtime, and fidelity overheads in QA workloads. Dense and irregular workloads often require long chains that increase physical-qubit usage, routing fragmentation, and hardware underutilization, reducing usable logical capacity despite the large number of physical qubits available on modern QA systems today \cite{macaskillsmith2026emberextensiblebenchmarksuite, gomeztejedor2026addressingminorembeddingproblemquantum, ayanzadeh2024skipper}.

Embedding also introduces significant runtime overhead because finding high-quality embeddings is computationally challenging. Determining the optimal embedding is NP-hard, forcing practical QA systems to rely on heuristic embedders such as \texttt{minorminer}~\cite{cai2014practical}. These heuristics may require substantial runtime and can fail even when feasible embeddings exist, requiring repeated attempts \cite{cai2014practical, gomeztejedor2026addressingminorembeddingproblemquantum, ayanzadeh2022equal}. While embedding latency can be amortized for static workloads, many practical applications, such as logistics and routing, continuously evolve. In such settings, workload changes require recomputing embeddings for updated Ising graphs, making embedding latency a critical bottleneck for time-sensitive and real-time applications.

Embedding additionally introduces tradeoffs between logical consistency and solution fidelity. If chain strength is too weak, chains are more likely to break during annealing when qubits within the same chain disagree at measurement time \cite{Raymond_2020, ayanzadeh2022equal}. Broken chains degrade logical consistency and require post-processing to recover logical outputs. In contrast, excessively strong chain couplings may distort the original optimization objective and reduce solution quality \cite{Raymond_2020, ayanzadeh2022equal, dwave_chain_strength_intro}. Thus, embedding directly shapes hardware utilization, execution fidelity, runtime behavior, and overall workload scalability on QAs.

Given its critical role in the quantum annealing pipeline, minor embedding has received considerable attention, including the recent Ember framework. Ember provides a benchmarking suite for comparing heuristic embedding algorithms across diverse graph instances and D-Wave hardware topologies \cite{macaskillsmith2026emberextensiblebenchmarksuite}. Using metrics such as embedding success rate, chain length, and runtime across six embedders, Ember provides a systematic view of algorithmic performance and a foundation for future embedding studies. While Ember focuses on embedding algorithms, our work takes a complementary systems and architecture perspective, characterizing embedding as a key bottleneck in the end-to-end quantum annealing pipeline.

\subsection{Goal of This Work}
This work characterizes embedding as a workload-dependent systems bottleneck in quantum annealing and examines its impact on end-to-end QA execution. We study how workload topology, graph structure, and embedding behavior affect physical-qubit utilization, runtime overheads, chain formation, and execution fidelity across QA systems.

\section{Methodology}

In this section, we describe the evaluation methodology used in this work.

\noindent
\textbf{Hardware Platform.}
We use the D-Wave Advantage\_system4.1 QA system, which employs the Pegasus topology and includes more than 4700 superconducting qubits and over 40000 couplers.
All experiments use the default forward annealing schedule, an annealing time of 20~$\mu$s, and 1000 samples per job~\cite{mcgeoch2021advantage,dwave2024topologies}.

\vspace{0.05in}
\noindent
\textbf{Software Platform.}
We use D-Wave Ocean SDK version 9.3.0 for workload generation, embedding, and QA execution.
Embedding is performed using the minorminer heuristic~\cite{D-Wave_Ocean_SDK} to map arbitrary logical graphs onto the QA hardware topology.
Graph workloads are generated using the NetworkX Python package.
Embedding experiments use a 1000-second timeout and multiple embedding attempts to reduce heuristic variability and embedding failures.

\vspace{0.05in}
\noindent
\textbf{Benchmarks.}
To study the impact of workload topology on embedding behavior, we evaluate six representative graph families: clique (SK model), complete bipartite, square grid, Erd\H{o}s-R\'enyi (ER), regular, and power-law or Barab\'asi--Albert (BA) graphs.
For each randomly generated graph, Ising coefficients are sampled uniformly from $[-1,+1]$.
We generate 10 random coefficient realizations for each workload configuration.
This workload suite is intentionally topology-centric rather than application-specific, enabling controlled characterization of how graph structure impacts embedding behavior and QA execution outcomes.

\vspace{0.05in}
\noindent
\textbf{Figures of Merit.}
We evaluate QA behavior using both embedding-level and execution-level metrics.
At the embedding level, we measure mean chain length and the 95th percentile chain length (p95 chain length) to characterize embedding overhead and long-chain behavior.
The 95th percentile metric captures heavy-tail effects where a small number of dominant chains disproportionately impact hardware utilization and execution reliability.

At the execution level, we measure chain-break ratio and solution quality.
Chain-break ratio measures the fraction of logical chains whose physical qubits disagree at the time of measurement. This metric captures logical inconsistency introduced during annealing.
To evaluate solution quality, we measure the energy gap between the best solution returned by the QA and a classical reference solution.
For the classical reference, we use brute-force enumeration for instances with $|V| \leq 22$ and simulated annealing for larger instances.
As a result, energy gaps for $|V|>22$ are measured against the best heuristic reference energy found rather than a certified global optimum~\cite{kirkpatrick1983optimization,dwave_neal}.

For chain-strength comparison, we first select a single practical baseline policy using a normalized multi-metric score that jointly balances the chain-break ratio, the energy gap, and the ratio of valid unbroken samples.
The full scoring rule is described in further detail in Section~\ref{sec:chain-strength}.
Together, these metrics directly connect workload topology and embedding structure to downstream QA execution fidelity, runtime behavior, and overall system-level performance across workloads.

\section{Effective Device Capacity}
\label{sec:embedding-capacity}

The effective capacity of a QA depends not only on the number of available physical qubits, but also on how efficiently a workload topology can be embedded onto the hardware graph.
In this section, we characterize how workload structure impacts the largest embeddable logical graph size, chain formation behavior, and physical-qubit utilization.
Our results show that effective QA capacity is fundamentally workload-dependent and can differ substantially across graph families on the same hardware platform.

To characterize effective device capacity, we first measure the largest logical graph that can be embedded for each workload topology.
For each graph family, we sweep its natural size parameter and report the corresponding number of logical variables, $|V|$.
We use binary search over candidate graph sizes to identify the largest embeddable instance for each topology.
Each candidate size is evaluated with a 500\,s embedding timeout.
Deterministic graph families use a single embedding attempt per size, while stochastic graph families use 10 random instances per size and require at least 8 successful embeddings to be considered reliably embeddable.

Different workload topologies use the most suitable embedding strategy available in existing prior work.
Complete graphs use the dedicated clique embedding routine~\cite{boothby2016fast}, complete bipartite graphs use the biclique embedding routine~\cite{boothby2020Pegasus}, and the remaining graph families use the minorminer heuristic embedder~\cite{cai2014practical}, respectively.

\begin{figure}[H]
    \centering
    \includegraphics[width=\columnwidth]{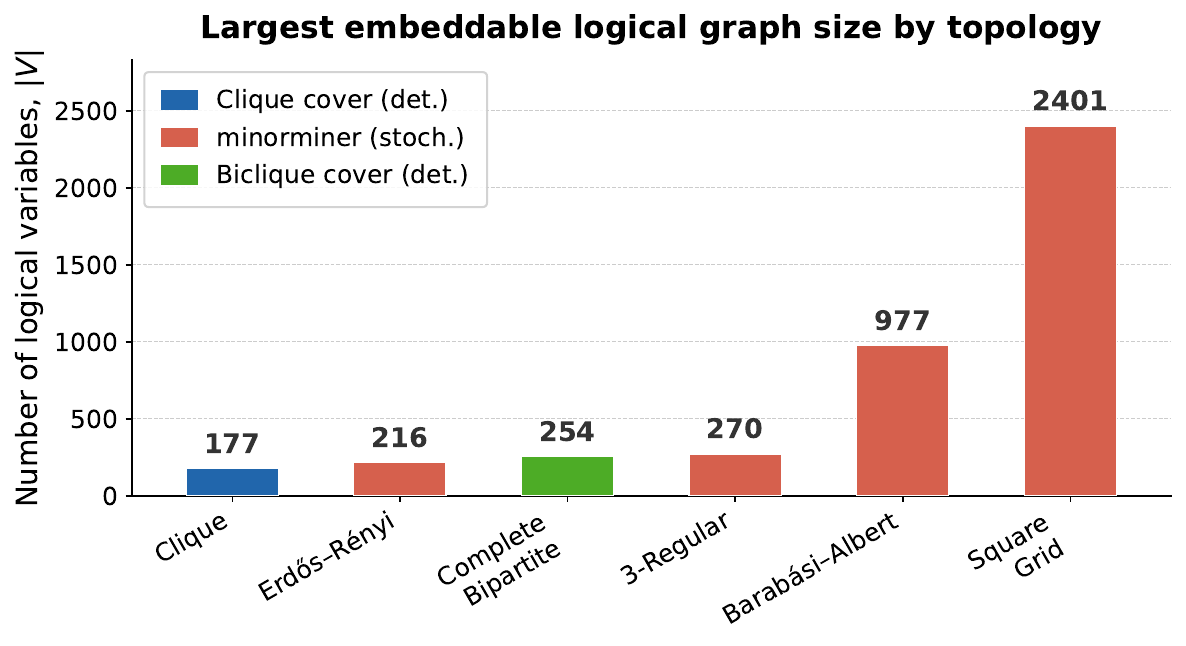}
    \caption{Largest embeddable logical graph size ($|V|$) per topology across three embedding methods.
    Complete bipartite $K_{127,127}$ yields $|V| = 254$, while the $49 \times 49$ square grid yields $|V|=2401$.}
    \label{fig:largest_embeddable_linear}
\end{figure}

Figure~\ref{fig:largest_embeddable_linear} shows that effective logical capacity strongly depends on workload topology.
Sparse local structures map most efficiently onto QA hardware.
For example, the square-grid topology reaches $49 \times 49$, corresponding to 2401 logical variables. Barab\'asi--Albert graphs also support relatively large problem instances, reaching up to 977 logical variables because they remain globally sparse despite the presence of high degree hubs.

In contrast, dense workloads reach the hardware limit much earlier.
Fully connected cliques support only 177 logical variables, while Erd\H{o}s--R\'enyi graphs support 216 logical variables. Complete bipartite graphs and 3-regular graphs fall between these extremes, reaching 254 and 270 logical variables, respectively.
These results show that the same QA hardware can support thousands of logical variables for sparse workloads but only a few hundred for dense or highly connected topologies.

\begin{figure}[t]
    \centering
    \includegraphics[width=\columnwidth, height=0.15\textheight]{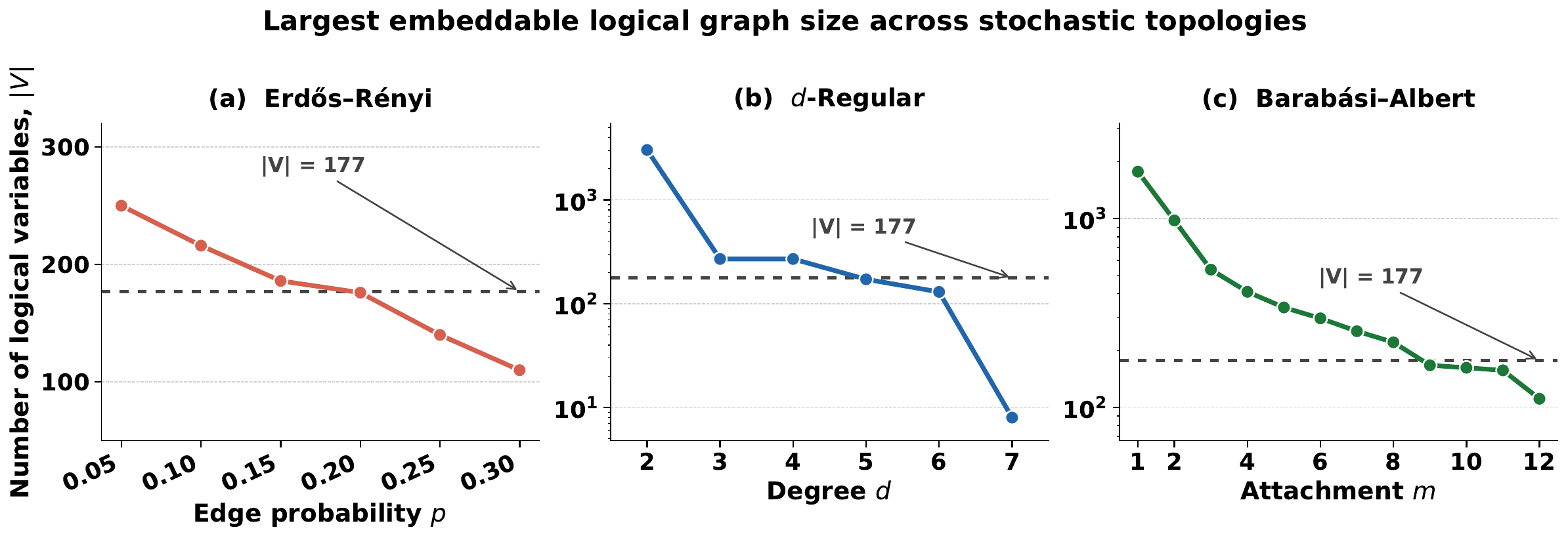}
    \caption{Largest embeddable logical graph size ($|V|$) versus connectivity parameter for three stochastic topologies:
    \Erdos edge probability $p$, $d$-Regular degree $d$, and Barab\'asi--Albert attachment parameter $m$.
    The dashed line marks the clique baseline ($|V|=177$).}
    \label{fig:stochastic_capacity_sweep}
\end{figure}

Figure~\ref{fig:stochastic_capacity_sweep} further shows that effective capacity decreases as workload connectivity increases.
For Erd\H{o}s--R\'enyi graphs, increasing edge probability $p$ introduces additional random interactions throughout the logical graph, increasing routing pressure on the hardware topology.
For regular graphs, increasing degree $d$ forces each logical variable to maintain more interactions, reducing the largest embeddable size toward the dense clique regime.
Similarly, increasing the attachment parameter $m$ in Barab\'asi--Albert graphs increases connectivity around hub variables and reduces embeddable capacity.
Across all workload families, increasing connectivity consistently substantially reduces the largest embeddable logical problem size.

An interesting trend appears in Figure~\ref{fig:stochastic_capacity_sweep}.
For Erd\H{o}s--R\'enyi, regular, and Barab\'asi--Albert graphs, increasing the topology parameter progressively reduces the largest embeddable logical size.
At sufficiently high connectivity, these workloads eventually fall below the clique reference line, despite cliques representing the densest possible logical topology.
For example, while the dedicated clique embedding supports 177 fully connected logical variables, highly connected stochastic workloads may support significantly fewer than 100 logical variables.

This behavior shows that effective QA capacity depends not only on workload topology, but also on the embedding strategy itself.
Clique and biclique workloads use topology-specific embedding methods optimized for those graph structures.
In contrast, Erd\H{o}s--R\'enyi, regular, and Barab\'asi--Albert graphs rely on the topology-agnostic minorminer heuristic.
As a result, achievable logical capacity can differ substantially between specialized embedding routines and general-purpose embedding heuristics.
These results suggest that topology-aware embedding methods can significantly improve effective QA capacity compared to topology-agnostic approaches.

To understand why such a large gap exists between logical and physical capacity, we next study chain formation behavior.
Table~\ref{tab:chain_stats_frontier} reports chain statistics at the largest embeddable logical size for each workload family.

Table~\ref{tab:chain_stats_frontier} shows that logical variables are often represented using multiple physical qubits, and that chain behavior strongly depends on workload topology.
Square-grid and regular graphs produce short chains, with mean chain lengths of only 1.35 and 2.01, respectively.
In contrast, dense and irregular workloads require substantially longer chains.
Clique, complete bipartite, and Erd\H{o}s--R\'enyi graphs require mean chain lengths of 16.67, 14.18, and 20.67, respectively.
Longer chains increase physical-qubit usage and are more difficult to maintain consistently during annealing, increasing the likelihood of chain breaks and reducing execution reliability.

Barab\'asi--Albert graphs exhibit a different behavior.
Although their average chain length remains moderate at 4.50, the maximum chain length reaches 66 physical qubits.
This indicates that a small number of hub variables dominate the embedding, requiring disproportionately long chains.
These results show that average chain length alone is insufficient to characterize embedding complexity since a few dominant chains can substantially impact overall embedding quality and execution fidelity.

However, chain overhead alone does not fully explain the gap between logical and physical capacity on QA hardware.
Even after embedding is completed successfully, many physical qubits remain unused across the hardware graph.

\begin{figure}[t]
    \centering
    \includegraphics[width=\columnwidth]{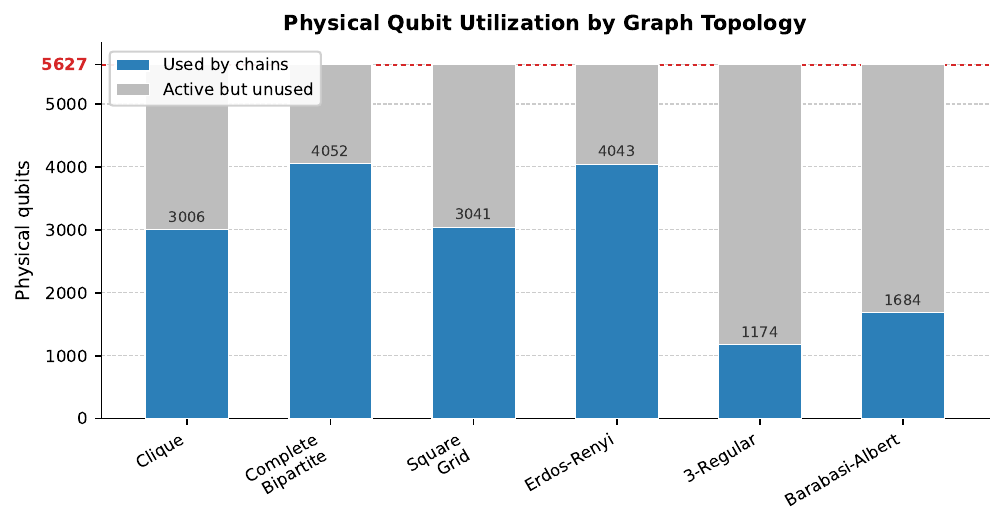}
    \caption{Physical-qubit utilization per topology, split between qubits used by chains and qubits active but unused after embedding.}
    \label{fig:used_unused_qubits}
\end{figure}

Figure~\ref{fig:used_unused_qubits} shows that a substantial fraction of physical qubits remain unused across several workload families.
These unused qubits cannot necessarily be repurposed to support additional logical variables because they may become isolated between chains and lack the couplers required to preserve logical interactions.
As a result, embedding overhead arises not only from representing logical variables using multiple physical qubits, but also from fragmentation of the remaining hardware graph.
Consequently, effective QA capacity depends jointly on chain overhead and on how efficiently the remaining hardware connectivity can be utilized after embedding the problem graph.

\begin{table}[h]
    \centering
    \caption{Chain statistics at the largest embeddable logical size.}
    \label{tab:chain_stats_frontier}
    \resizebox{\columnwidth}{!}{
    \begin{tabular}{lrrrr}
    \toprule
    Workload topology & $|V|$ & Mean chain & Max chain & Chain var. \\
    \midrule
    Clique              & 177  & 16.67 & 17 & 0.22 \\
    Complete bipartite  & 254  & 14.18 & 15 & 0.15 \\
    Square grid         & 2401 & 1.35  & 4  & 0.32 \\
    Erd\H{o}s--R\'enyi  & 216  & 20.67 & 44 & 58.23 \\
    Random regular      & 270  & 2.01  & 6  & 0.88 \\
    Barab\'asi--Albert  & 977  & 4.50  & 66 & 51.04 \\
    \bottomrule
    \end{tabular}
    }
    \end{table}

\begin{figure}[t]
    \centering
    \includegraphics[width=\columnwidth]{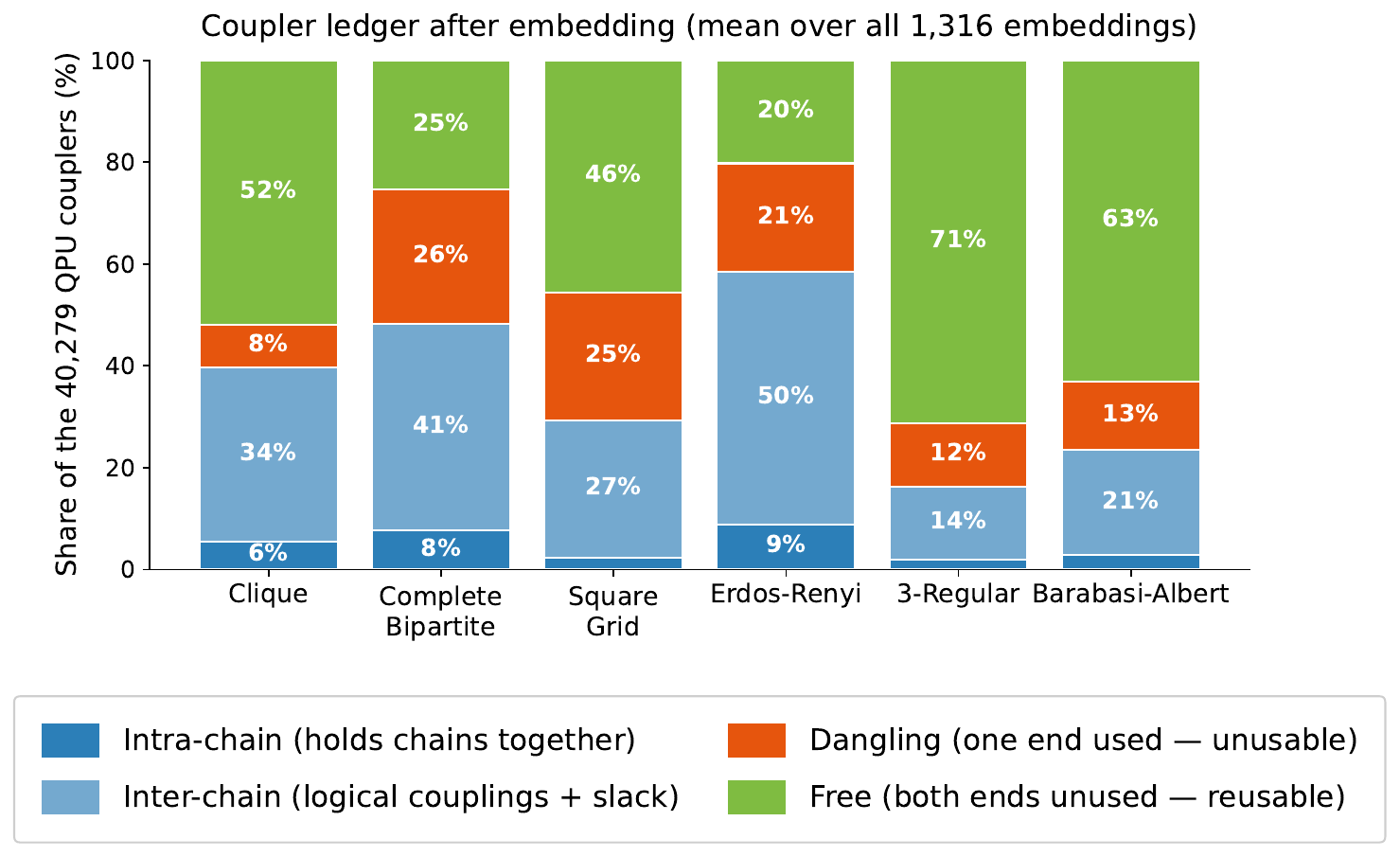}
    \caption{Physical-coupler utilization after embedding across workload topologies, averaged over 1,316 embeddings. Dangling couplers have one occupied and one unused endpoint and therefore cannot support an additional embedding.}
    \label{fig:coupler_fragmentation}
\end{figure}

To better understand this hardware underutilization, we next characterize how the unused qubits and couplers are distributed after embedding. 
Across 1,316 embeddings, we find that the
unused qubits are generally not divided into small isolated regions. 
The largest connected component contains between 95.1\% and 99.9\% of the unused qubits across workload families, showing that most remaining qubits are still physically connected.
However, this connectivity does not imply that the remaining hardware can be efficiently reused. 
Embedding also consumes couplers required to maintain interactions between
physical qubits. 
As shown in Fig.~\ref{fig:coupler_fragmentation},
between 8.4\% and 26.4\% of all physical couplers become dangling after embedding, where one endpoint is occupied and
the other remains unused. 
As a result, only approximately 20\% to 71\% of the couplers remain fully available, depending on workload topology. Despite thousands of unused physical
qubits, the remaining hardware supports clique embeddings of only $K_{16}$ to $K_{64}$. 
These results show that hardware
fragmentation is driven primarily by the loss of usable connectivity rather than by isolated unused qubits. 
 Physical-qubit utilization alone can substantially overestimate
the capacity available for additional workloads.

\section{Embedding Runtime and Reliability}

Embedding impacts not only effective QA capacity, but also overall system performance and execution reliability.
Before annealing can begin, the logical workload must first be transformed into a hardware-compatible representation, placing embedding directly on the critical execution path.
As a result, embedding latency contributes directly to end-to-end time-to-solution.
Moreover, because practical embedders are heuristic, embedding runtime can vary substantially across workloads and embedding attempts, and embedding may fail even when feasible mappings exist.
These effects become particularly important for dynamic and time-sensitive applications where workloads evolve over time and embeddings must be recomputed repeatedly.

To characterize the runtime and reliability of embeddings, we measure embedding latency and success rate across diverse workload topologies.
Table~\ref{tab:embedding_runtime} reports the minimum successful embedding time, amortized mean embedding time, and embedding success rate for representative workload instances near the embedding frontier.

\begin{table}[]
\centering
\caption{Embedding runtime by workload topology.
Minimum runtime is the fastest successful trial, while amortized mean penalizes failed trials using the full timeout.}
\label{tab:embedding_runtime}
\resizebox{\columnwidth}{!}{
\begin{tabular}{l l r r r}
\toprule
Workload topology & Size & Min. time & Amort. mean & Success \\
\midrule
Clique             & $165$  & 75.0s  & 217.7s & 100\% \\
Complete bipartite & $238$  & 125.5s & 462.9s & 100\% \\
Square grid        & $1369$ & 12.4s  & 171.4s & 90\%  \\
Erd\H{o}s--R\'enyi & $194$  & 26.4s  & 94.6s  & 100\% \\
Random regular     & $450$  & 2.9s   & 6.5s   & 100\% \\
Barab\'asi--Albert & $512$  & 19.6s  & 43.1s  & 100\% \\
\bottomrule
\end{tabular}
}
\end{table}

\begin{figure}[H]
    \centering
    \includegraphics[width=\columnwidth]{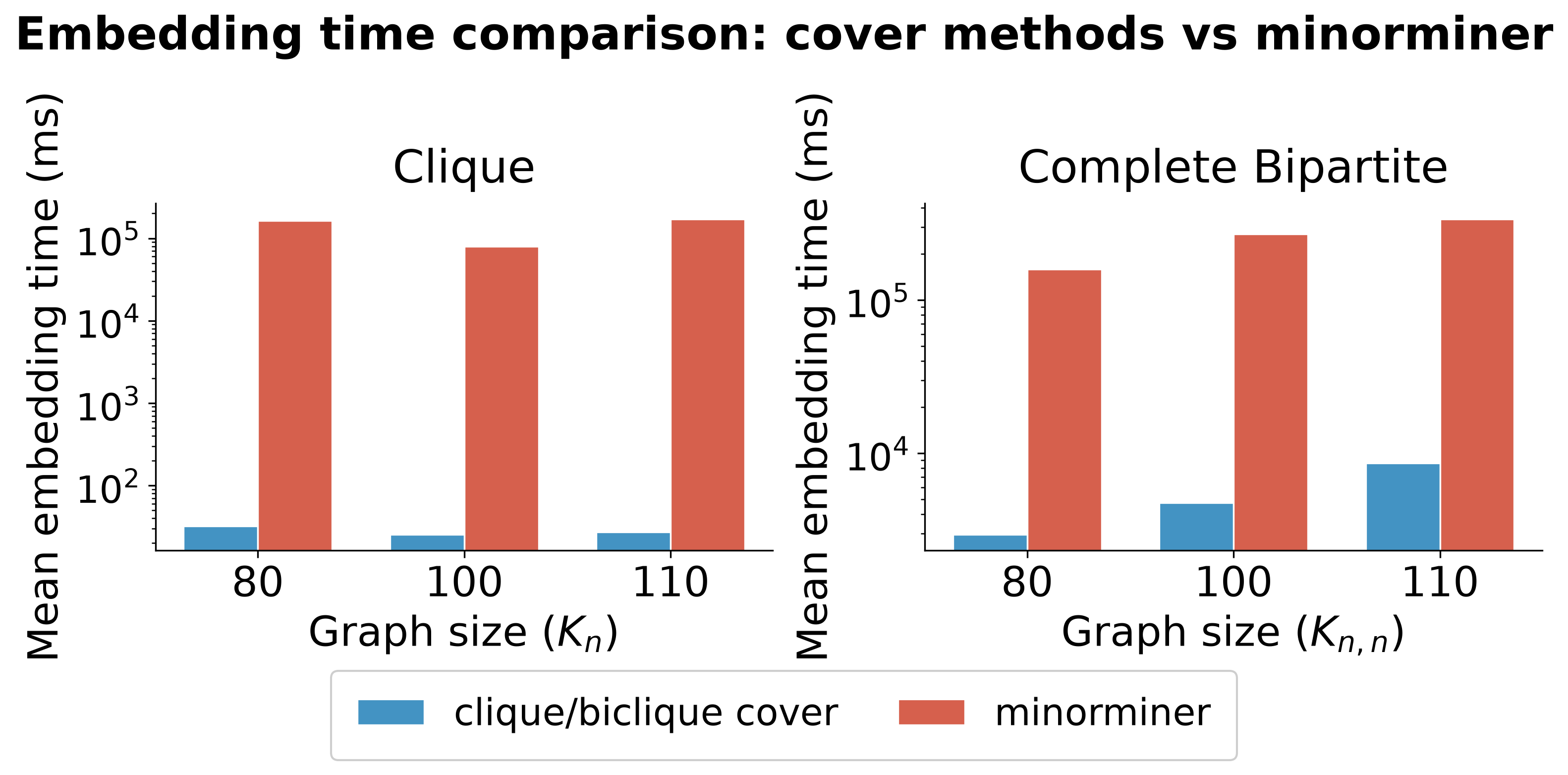}
    \caption{Mean embedding runtime (log scale) for clique and complete bipartite workloads comparing topology-specific embedding routines against the topology-agnostic method.}
    \label{fig:embedding-time-cover-vs-minorminer}
\end{figure}

Table~\ref{tab:embedding_runtime} shows that embedding runtime strongly depends on workload topology and embedding strategy.
Clique and complete bipartite workloads require the longest embedding times despite their smaller logical sizes.
In contrast, regular graphs embed substantially faster, while Erd\H{o}s--R\'enyi and Barab\'asi--Albert workloads fall between these two extremes.
These results show that embedding latency is not determined solely by logical graph size.
Instead, runtime depends on how difficult it is to route the logical graph through the sparse hardware topology and on whether the embedding method can exploit workload structure efficiently.

The table also shows substantial variability between minimum embedding time and amortized embedding time.
This gap reflects the stochastic nature of heuristic embedding and the cost of failed embedding attempts.
For example, the square-grid topology achieves a minimum embedding time of only 12.4\,s, yet its amortized runtime increases to 171.4\,s due to failed embedding attempts and embedding variability.
These results show that embedding reliability is an important systems concern because embedding failures directly increase runtime and may prevent workload execution entirely.

Figure~\ref{fig:embedding-time-cover-vs-minorminer} further shows that embedding runtime depends heavily on the embedding method itself.
For clique and complete bipartite workloads, topology-specific embedding routines are orders of magnitude faster than the topology-agnostic minorminer heuristic at the same graph sizes.
This highlights an important distinction between specialized and general-purpose embedding approaches.
Topology-specific embedding methods can directly exploit known graph structure and avoid much of the expensive search process required by topology-agnostic heuristics.

In practice, embedding cost also depends on whether embeddings can be reused across executions.
For some workload classes, such as clique workloads, the same embedding can often be reused across different applications because the logical topology remains unchanged and only Ising coefficients differ.
For example, applications such as TSP, SAT, and compressive sensing may all reuse the same clique embedding once it has been generated for a target hardware graph.
In such settings, embedding latency can be amortized across many executions.

However, many practical applications involve dynamically changing workload topologies.
For example, applications such as logistics, routing, scheduling, and adaptive network optimization may continuously modify the underlying logical graph during execution.
In these settings, embeddings must be recomputed repeatedly at runtime, making embedding latency and embedding reliability part of the online execution path.
Importantly, these dynamic workloads often correspond to the most challenging embedding regimes, including Erd\H{o}s--R\'enyi, regular, and Barab\'asi--Albert graphs, where topology-specific embedding methods are unavailable and general-purpose heuristics must be used.
As a result, embedding failures, runtime variability, and long embedding latency can directly limit responsiveness and scalability in real-time QA applications.

These results highlight that embedding overhead cannot be evaluated independently from workload structure.
Even workloads with comparable logical sizes may exhibit substantially different embedding runtime and reliability due to differences in graph topology and routing complexity.
Consequently, workload topology becomes a first-order factor in determining practical QA usability and end-to-end execution efficiency.

Overall, these results clearly show that embedding is not merely a preprocessing step, but a major classical systems bottleneck in practical quantum annealing.
A workload may fit on the QPU and achieve acceptable chain behavior, yet still require seconds or even minutes of classical preprocessing before a short annealing execution can begin.
Consequently, QA workload characterization must consider embedding runtime and reliability alongside logical capacity, chain statistics, execution fidelity, and overall system performance.

\section{Chain Strength and Execution Fidelity}
\label{sec:chain-strength}

After embedding determines whether a workload can execute on QA hardware, chain strength largely determines execution reliability during annealing.
Different workload topologies produce different chain structures and chain lengths, causing chain-strength sensitivity to vary substantially across workloads.
In this section, we characterize how chain-strength policies impact execution reliability and solution quality across diverse QA workloads.

Chain strength controls the ferromagnetic coupling applied within each chain to encourage all physical qubits representing the same logical variable to return a consistent value~\cite{dwave_chain_strength_intro}.
If chain strength is too weak, chains are more likely to break during sampling, producing inconsistent logical assignments.
In contrast, excessively strong chain couplings may distort the target energy landscape and negatively affect solution quality.
As a result, chain-strength selection introduces a tradeoff between chain reliability and optimization fidelity.

We evaluate chain strength using two metrics.
The chain-break ratio measures the fraction of logical chains broken in returned QPU samples.
The energy gap measures the difference between the best unembedded QPU energy and the classical reference energy.
Lower values are better for both metrics, although minimizing one metric does not necessarily minimize the other.

We compare several chain-strength policies, including the Ocean default policy, uniform torque compensation (UTC), scaled maximum-coupling policies, and a mean-coupling policy.
UTC estimates chain strength using workload connectivity and quadratic biases~\cite{dwave_utc}, while scaled policies define chain strength relative to the problem bias range~\cite{dwave_scaled}.
The scaled maximum-coupling policies set chain strength proportional to $\alpha \max_{i,j}|J_{ij}|$, where $\alpha \in \{0.5,1.0,1.5,2.0,3.0\}$.
The mean-coupling policy sets chain strength proportional to $\mathrm{mean}_{i,j}|J_{ij}|$.
For each graph family and policy, we execute the embedded Ising workloads on the QPU, unembed the returned samples, and compute the mean chain-break ratio and mean energy gap.

\begin{figure}[h]
    \centering
    \includegraphics[width=\columnwidth,height=0.22\textheight]
   {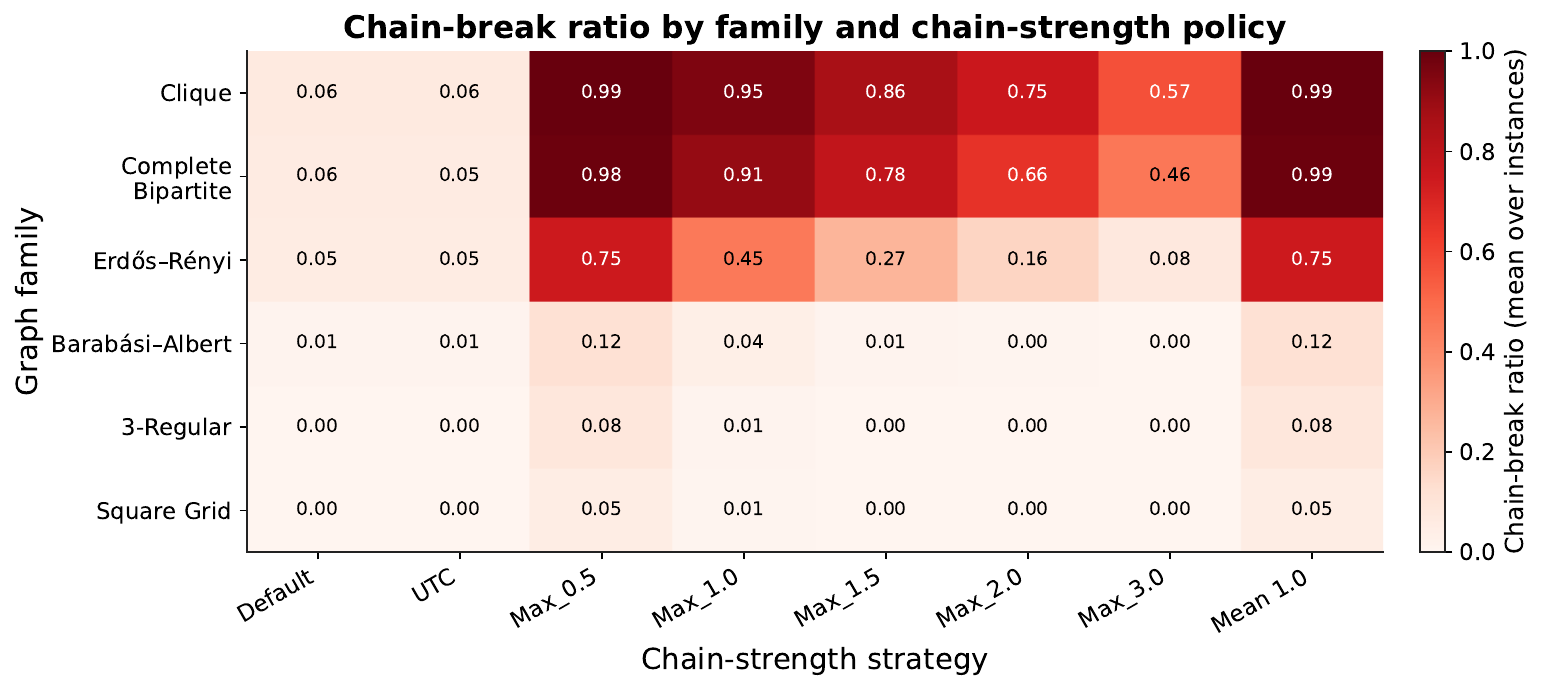}
    \caption{\textbf{Chain-break ratio across workload topologies and chain-strength policies.}
    Weak or mismatched chain-strength policies cause severe reliability degradation for dense workloads, while default and UTC maintain consistently low chain-break ratios across all evaluated topologies.}
    \label{fig:chain-break-heatmap}
\end{figure}

Figure~\ref{fig:chain-break-heatmap} shows that chain-strength policy strongly affects execution reliability.
Default and UTC maintain chain-break ratios at or below approximately $0.06$ across all evaluated workload families.
In contrast, clique and complete bipartite workloads exhibit chain-break ratios approaching one under weak or poorly matched policies.
Erd\H{o}s--R\'enyi workloads show intermediate sensitivity, while Barab\'asi--Albert, regular, and square-grid workloads remain stable because their embeddings use shorter chains.

\begin{figure}[h]
    \centering
\includegraphics[width=\columnwidth,height=0.22\textheight]{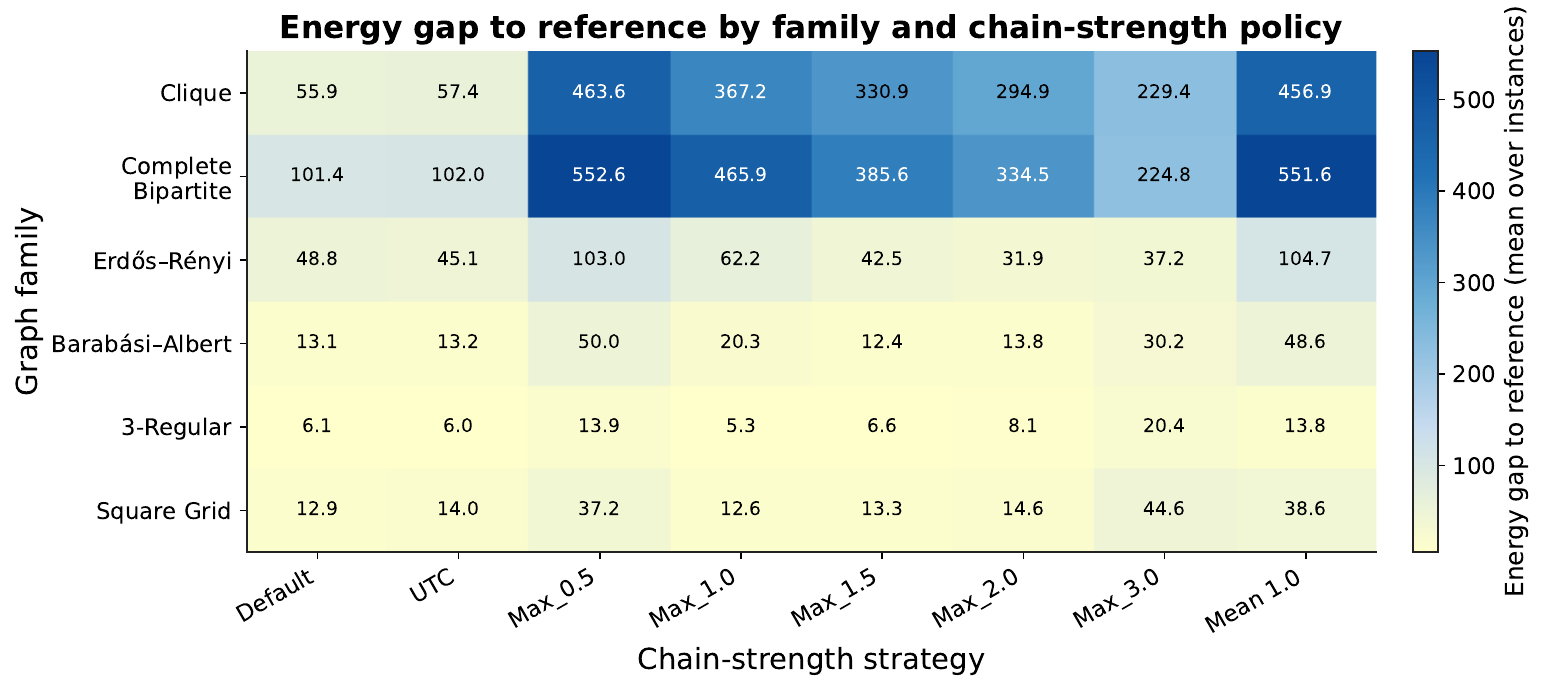}
    \caption{\textbf{Energy gap across workload topologies and chain-strength policies.}
    Policies minimizing chain-break ratio do not necessarily minimize energy gap, showing that chain-strength selection introduces a tradeoff between execution reliability and optimization fidelity.}
    \label{fig:energy-gap-heatmap}
\end{figure}

Figure~\ref{fig:energy-gap-heatmap} shows that chain-strength selection also substantially impacts solution quality across the evaluated workloads.
For clique and complete bipartite workloads, several non-default policies produce significantly larger energy gaps than default or UTC.
For Erd\H{o}s--R\'enyi and sparse workloads, the energy gap varies less dramatically, although the policy minimizing energy gap can still differ from the policy minimizing chain-break ratio.
These results show that chain strength cannot be evaluated using chain reliability alone.

To select a practical baseline policy for the remaining characterization, we use a balanced score combining chain-break ratio, energy gap, and the fraction of valid unbroken samples.
The score normalizes the metrics onto a common scale to balance execution reliability and solution quality.
It is used only to select a practical baseline for this study rather than define a universally optimal chain-strength policy.

\begin{table}[H]
\centering
\caption{\textbf{Best-performing chain-strength policies by metric and selected baseline policy.}
Policies minimizing chain-break ratio and energy gap often differ across workload topologies, highlighting the multi-objective nature of chain-strength selection.}
\label{tab:chain-strength-policy}
\scriptsize
\setlength{\tabcolsep}{3pt}
\renewcommand{\arraystretch}{1.12}
\begin{tabular}{l c c c}
\toprule
Workload topology & Lowest break & Lowest gap & Selected baseline \\
\midrule
Barab\'asi--Albert   & $3.0\times\max|J|$ & $1.5\times\max|J|$ & default \\
Clique               & default            & default            & default \\
Complete bipartite   & UTC                & default            & default \\
Erd\H{o}s--R\'enyi   & default            & $2.0\times\max|J|$ & default \\
Regular              & $3.0\times\max|J|$ & $1.0\times\max|J|$ & default \\
Square grid          & $3.0\times\max|J|$ & $1.0\times\max|J|$ & default \\
\bottomrule
\end{tabular}
\vspace{-0.5em}
\end{table}

Table~\ref{tab:chain-strength-policy} further shows that chain-strength selection is inherently multi-objective.
Several workload topologies exhibit different policies for minimizing chain-break ratio and minimizing energy gap.
Although workload-specific tuning can improve individual metrics, no single tuned policy consistently improves all metrics across workloads.
Overall, default and UTC provide the most robust cross-workload behavior, maintaining low chain-break ratios while avoiding the large energy-gap penalties observed under several alternative policies.

Overall, these results show that chain strength is not merely a tuning parameter, but a workload-dependent execution reliability mechanism.
Different workload topologies produce different chain structures and reliability behaviors, causing the optimal chain-strength policy to vary across workloads.
Consequently, future QA runtime systems may require topology-aware or adaptive chain-strength selection rather than relying on static global policies.

\section{Embedding Quality and QPU Fidelity}
\label{sec_embedding_prediction}

In this section, we investigate whether embedding quality can predict downstream QA execution behavior before annealing begins.
If embedding-side metrics strongly correlate with QPU outcomes, embedding quality could be estimated immediately after embedding without consuming QPU resources.

We study two execution outcomes: chain-break ratio and energy gap.
Chain-break ratio measures how frequently physical chains fail to represent valid logical variables, while energy gap measures the difference between the best QPU solution and the classical reference energy.
For each embedding, ChainForge records seven embedding-side metrics available prior to QPU execution: expansion ratio, physical qubits used, mean chain length, maximum chain length, p95 chain length, chain-length variance, and chain-length coefficient of variation.
The prediction dataset contains 927 embeddings across six workload families, split into 741 training embeddings and 186 held-out test embeddings.

The key metric is expansion ratio,
\[
\rho = \frac{|V|}{Q_{\mathrm{phys}}},
\]
where $|V|$ is the number of logical variables and $Q_{\mathrm{phys}}$ is the number of physical qubits used by the embedding.
A larger expansion ratio corresponds to a more compact embedding that uses relatively fewer physical qubits per logical variable.

For each embedding metric, we evaluate two prediction targets.
Break $R^2$ measures how well the metric predicts chain-break ratio, while Gap $R^2$ measures how well the metric predicts energy gap.
We additionally report rank correlation for both outcomes.

\begin{table}[t]
\centering
\caption{\textbf{Single-metric prediction strength for QA execution outcomes.}
Expansion ratio is the strongest individual predictor of both chain reliability and solution quality.}
\label{tab_predictor_ranking}
\scriptsize
\setlength{\tabcolsep}{3pt}
\renewcommand{\arraystretch}{1.12}
\begin{tabular}{l r r r r}
\toprule
Embedding metric & Break $R^2$ & Gap $R^2$ & Break corr. & Gap corr. \\
\midrule
Expansion ratio & 0.620 & 0.310 & -0.865 & -0.776 \\
Mean chain length & 0.609 & 0.221 & 0.865 & 0.776 \\
Physical qubits used & 0.477 & 0.240 & 0.600 & 0.791 \\
Chain-length CV & 0.442 & 0.301 & -0.608 & -0.614 \\
P95 chain length & 0.310 & 0.043 & 0.850 & 0.766 \\
Max chain length & 0.284 & 0.088 & 0.779 & 0.740 \\
Chain-length variance & 0.030 & 0.025 & 0.395 & 0.304 \\
\bottomrule
\end{tabular}
\vspace{-0.5em}
\end{table}

Table~\ref{tab_predictor_ranking} shows that expansion ratio is the strongest single predictor for both execution outcomes.
By itself, expansion ratio explains 62.0\% of the variation in chain-break ratio and 31.0\% of the variation in energy gap.
The negative correlation indicates that compact embeddings tend to produce fewer broken chains and smaller energy gaps.
Mean chain length provides similarly strong predictive power for chain breaks, although it is weaker for energy gap.
These results show that embedding compactness and chain structure jointly shape downstream QA reliability and solution quality.

\begin{figure}[b]
    \centering
    \includegraphics[width=\columnwidth]{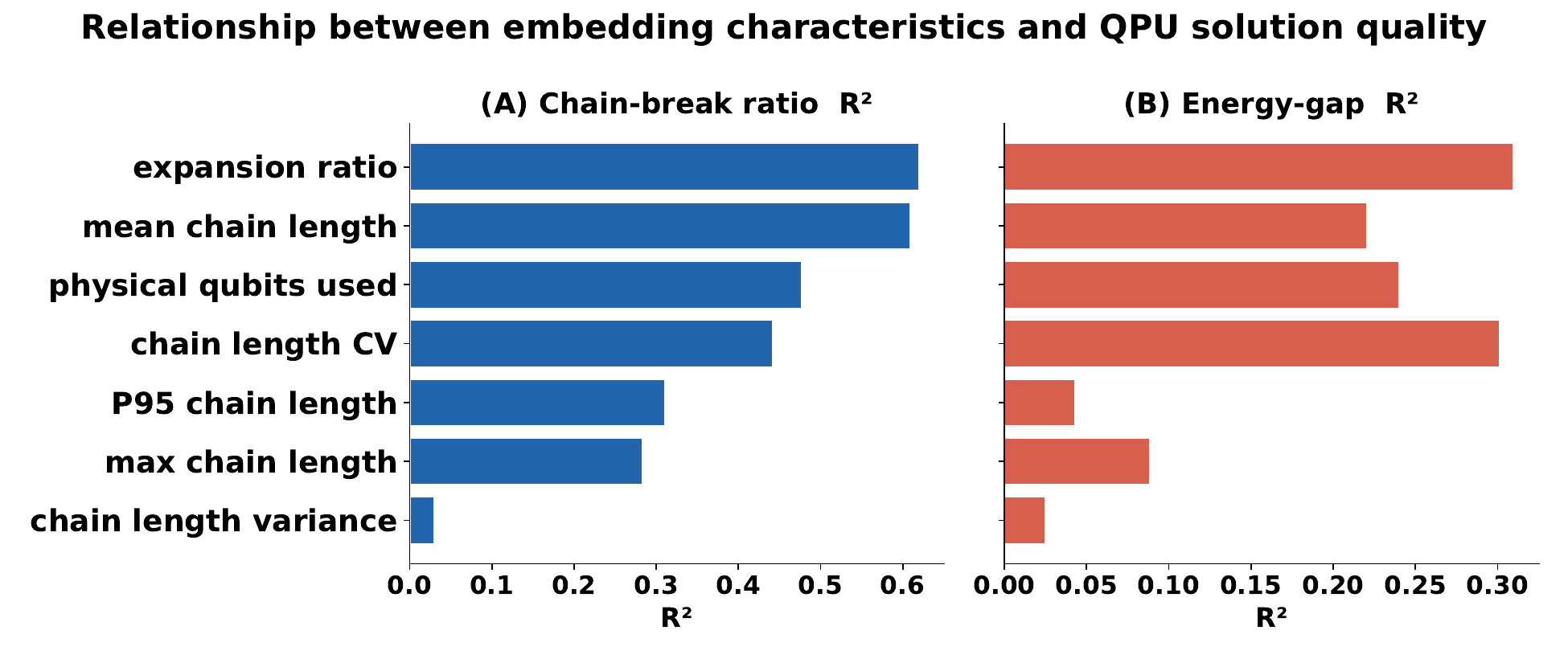}
    \caption{\textbf{Prediction strength of individual embedding metrics.}
   Expansion ratio and mean chain length substantially outperform other embedding metrics in predicting downstream QA behavior, providing stronger indicators of execution quality and overall system-level performance.
}
    \label{fig_single_predictor_r2}
\end{figure}

Figure~\ref{fig_single_predictor_r2} further shows that embedding compactness metrics consistently outperform raw chain-distribution statistics.
Importantly, these metrics are available immediately after embedding and therefore require no QPU execution overhead.
This enables embedding quality to be estimated before annealing begins.

We next evaluate whether combining multiple embedding metrics improves prediction accuracy.
We use a reduced four-metric linear model consisting of maximum chain length, chain-length variance, chain-length coefficient of variation, and expansion ratio.
This model achieves a held-out $R^2$ of 0.896 for predicting chain-break ratio, while the training $R^2$ remains close at 0.908, indicating good generalization across unseen embeddings.

\begin{figure}[h]
    \centering
    \includegraphics[width=\columnwidth]{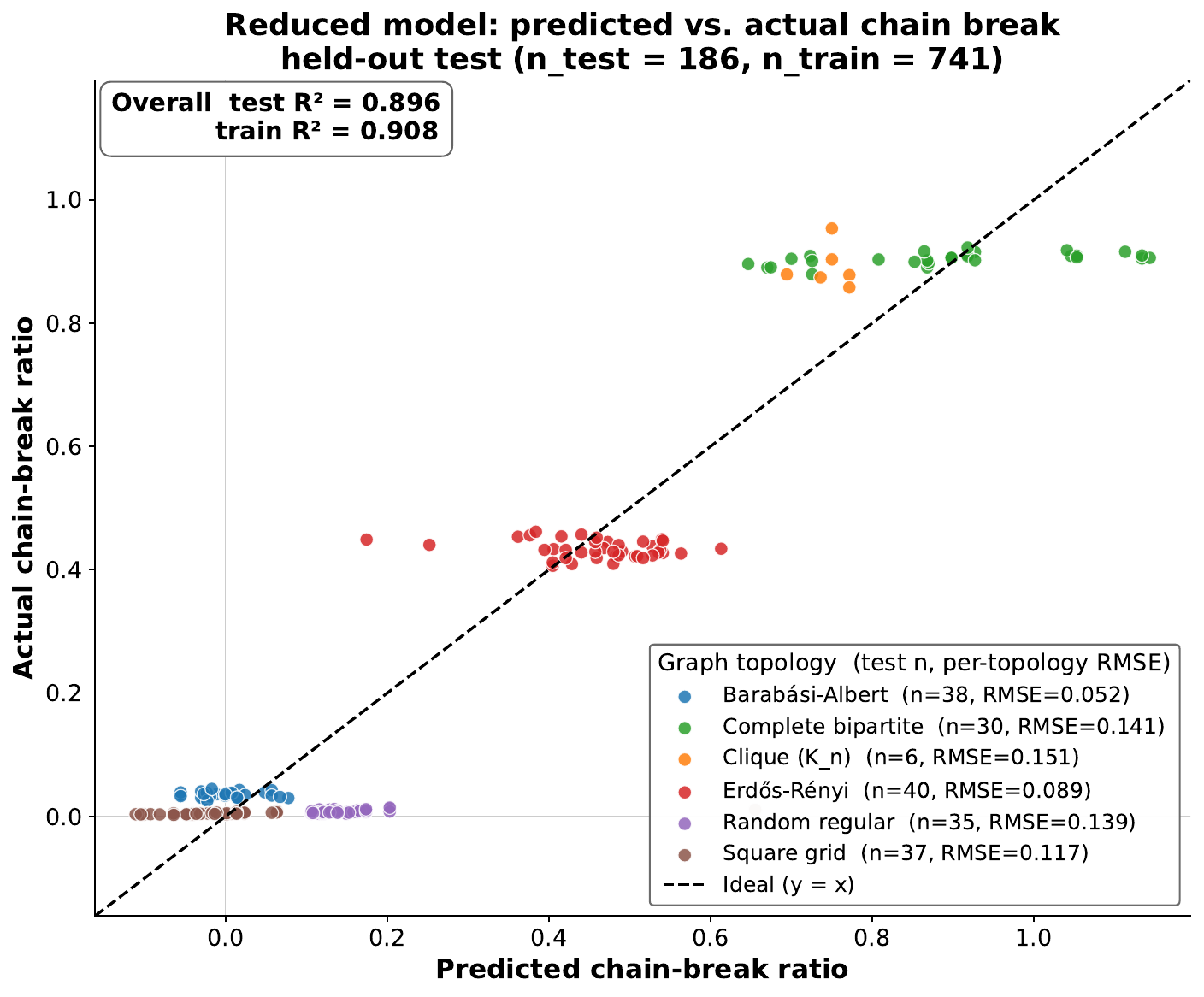}
    \caption{\textbf{Predicted versus measured chain-break ratio on the held-out test set.}
    Embedding-side metrics strongly predict downstream QA execution reliability across diverse workload families.}
    \label{fig_predicted_vs_actual}
\end{figure}

Figure~\ref{fig_predicted_vs_actual} shows that the reduced model captures the dominant chain-break trends across all evaluated workload families.
Although prediction accuracy is not perfect for every embedding, the model is sufficiently accurate for workload characterization and embedding screening.
A QA runtime system could therefore estimate embedding quality immediately after embedding and identify unreliable embeddings before consuming QPU resources.

Overall, these results show that embedding quality can be characterized using low-cost embedding-side metrics prior to QPU execution.
This enables QA runtime systems to estimate execution reliability immediately after embedding, potentially avoiding unreliable embeddings before consuming QPU resources.
More broadly, embedding-side metrics provide lightweight runtime signals for workload screening, scheduling, filtering, and selecting candidate embeddings in future QA systems.

\section{Related Work}

\vspace{0.05in}
\noindent
\textbf{QA Hardware and Minor Embedding.}
The scalability of QAs is closely tied to hardware connectivity and embedding efficiency in large-scale optimization workloads.
Foundational work on minor embedding first introduced connected physical-qubit chains to represent individual logical variables on sparse QA hardware~\cite{choi2008minorembeddingadiabaticquantumcomputation}.
D-Wave architectures have steadily evolved from Chimera to Pegasus and then Zephyr to increase qubit connectivity and better support larger workloads~\cite{boothby2020Pegasus,boothby2021zephyr}.
Prior work has also proposed several topology-specific embedding methods, including clique and biclique embeddings, for dense graph families~\cite{boothby2016Chimera}.
Other studies have further analyzed embedding on imperfect hardware graphs with missing qubits or broken couplers~\cite{lobe2021minor,klymko2014adiabatic}.
Together, these efforts establish hardware connectivity as a fundamental practical constraint on the size and structure of logical problems realizable on a given target QA processor. Nominal qubit count alone, therefore, does not fully determine the QA capacity. 
The interaction between the workload topology and the hardware graph structure ultimately governs both embedding overhead and the achievable logical problem size.

\vspace{0.05in}
\noindent
\textbf{Embedding Algorithms and Benchmarks.}
Because optimal embedding is computationally intractable, practical QA workflows rely on heuristic embedding methods.
Minorminer is among the most widely used heuristics for finding graph minors on D-Wave hardware~\cite{cai2014practical}.
Other approaches use integer programming and graph-based optimization techniques to improve embedding quality for selected workloads~\cite{bernal2020integer,goodrich2018optimizing}.
Recent work has also explored embedding benchmarks and evaluation frameworks.
Gomez-Tejedor et al. compare multiple embedding methods across workloads and hardware topologies~\cite{gomeztejedor2026addressingminorembeddingproblemquantum}.
Ember further provides an extensible benchmark suite for systematically evaluating embedding algorithms across different graph structures and hardware topologies, with emphasis on metrics such as embedding success, chain characteristics, and embedding runtime~\cite{macaskillsmith2026emberextensiblebenchmarksuite}.

\vspace{0.05in}
\noindent
\textbf{Chain Strength and QA Fidelity.}
Prior work has studied the relationship between chain strength, chain breaking, and QA execution fidelity.
Existing studies have explored chain-break behavior, hardware bias, and parameter-selection techniques for improving QA solution quality~\cite{grant2022benchmarking,barbosa2021optimizing,ayanzadeh2022equal}.
D-Wave Ocean additionally provides practical chain-strength policies such as scaled chain strength and uniform torque compensation~\cite{dwave_scaled,dwave_utc}.
These studies show that embedding parameters can significantly affect execution reliability and solution quality, making chain management an important component of practical QA execution.

\vspace{0.05in}
\noindent
\textbf{QA Applications and Workloads.}
Many optimization problems can be formulated as Ising workloads, including graph coloring, SAT, routing, scheduling, portfolio optimization, traffic control, and cryptographic search~\cite{lucas2014ising,elsokkary2017financial,feld2019hybrid,hu2020quantum,inoue2021traffic,ayanzadeh2019quantum,ayanzadeh2020ensemble,ayanzadeh2018solving}.
Most prior QA application studies focus on solution quality or time-to-solution for specific application domains~\cite{McGeoch2020169,ayanzadeh2020leveraging,ayanzadeh2022quantum,Ayanzadeh2020Reinforcement}.
In these studies, embedding is typically treated as part of the execution pipeline rather than as the primary target of workload characterization.

\vspace{0.05in}
Overall, prior work has substantially advanced QA hardware, embedding algorithms, chain-strength policies, and application-level demonstrations.
However, existing embedding-benchmarking efforts \cite{gomeztejedor2026addressingminorembeddingproblemquantum,macaskillsmith2026emberextensiblebenchmarksuite} focus primarily on embedding-centric metrics such as chain length, embedding success rate, and embedding runtime, which do not fully capture downstream QA behavior, including chain-break rates, execution reliability, and final solution quality after unembedding.
These evaluations also provide limited workload diversity: several studies primarily focus on Erd\H{o}s-R\'enyi random graphs and explicitly identify broader graph-family evaluation as an open direction \cite{gomeztejedor2026addressingminorembeddingproblemquantum}, despite different workload topologies imposing fundamentally different demands on QA connectivity and chain formation.
This work complements these efforts by characterizing embedding as a workload-dependent systems bottleneck that shapes effective capacity, runtime behavior, and downstream QA execution fidelity, thereby providing the workload-centric characterization that prior evaluations have not yet systematically undertaken.
This perspective shifts embedding from a purely compiler-level preprocessing task to a important central systems abstraction governing practical QA scalability and execution behavior.

\newpage
\section{Discussion and Conclusion}

This work shows that embedding fundamentally defines the effective architecture of quantum annealers.
Although modern QAs contain thousands of physical qubits, practical logical capacity is determined by workload topology, embedding efficiency, and hardware connectivity rather than nominal qubit count alone.
Across the evaluated workload families, we observe substantial differences in embeddable logical capacity, chain behavior, runtime overheads, and execution fidelity on the same hardware platform.
These results show that embedding is not merely a preprocessing stage, but a dominant systems bottleneck governing practical QA scalability and usability.

A key observation is that topology-specific embedding methods substantially outperform topology-agnostic heuristics across multiple dimensions.
Specialized embedding routines achieve higher logical capacity, lower runtime overheads, and more reliable embeddings than general-purpose heuristics such as minorminer.
In several cases, stochastic workloads using topology-agnostic heuristics support fewer logical variables than dense clique workloads despite lower logical connectivity.
These observations suggest that future QA software stacks may require workload-specialized embedding pipelines rather than relying solely on general-purpose heuristics.
Embedding should therefore be treated as an integral part of runtime and system design, rather than as an offline preprocessing step.

Our results also expose an important hardware design implication.
Even when workloads are successfully embedded, substantial portions of the hardware remain unusable because embedding consumes connectivity needed to route logical interactions.
Consequently, increasing physical-qubit count does not translate proportionally into greater usable logical capacity.
Future QA architectures should therefore consider not only the number and degree of physical qubits, but also how efficiently their connectivity supports representative workloads.
This motivates tighter hardware-software co-design, where hardware topology and embedding mechanisms are jointly optimized for targeted workload families.

More broadly, this work positions embedding as a first-class systems abstraction for quantum annealing.
Workload topology determines how logical problems consume physical resources, while embedding translates these requirements into capacity, runtime, and fidelity costs.
This perspective creates opportunities for workload-aware embedding frameworks, topology-specialized methods, embedding-aware runtime policies, and hardware topologies designed around representative workload classes.
Ultimately, scaling future QAs requires not only more physical qubits, but also the ability to efficiently map and execute useful logical workloads.

\section*{Acknowledgments}
This work was supported in part by the National Science Foundation (NSF) under Grant No.~2544544.
We sincerely thank D-Wave for providing access to their QAs.
AI-based language tools were used solely for proofreading and for improving the overall clarity and readability of the manuscript.

\newpage
\bibliographystyle{IEEEtranS}
\bibliography{refs}

\end{document}